# A BEOL Compatible, 2-Terminals, Ferroelectric Analog Non-Volatile Memory

Laura Bégon-Lours
*IBM Research GmbH*
Rüschlikon, Switzerland
ORCID 0000-0003-2520-3317
lbe@zurich.ibm.com

Mattia Halter
*IBM Research GmbH*
Rüschlikon, Switzerland
*ETH Zürich*
Zürich, Switzerland
ORCID 0000-0001-8468-9105

Diana Dávila Pineda
*IBM Research GmbH*
Rüschlikon, Switzerland
dda@zurich.ibm.com

Youri Popoff
*IBM Research GmbH*
Rüschlikon, Switzerland
*ETH Zürich*
Zürich, Switzerland
ypo@zurich.ibm.com

Valeria Bragaglia
*IBM Research GmbH*
Rüschlikon, Switzerland
ORCID 0000-0003-0636-4211

Antonio La Porta
*IBM Research GmbH*
Rüschlikon, Switzerland
alp@zurich.ibm.com

Daniel Jubin
*IBM Research GmbH*
Rüschlikon, Switzerland
dju@zurich.ibm.com

Jean Fompeyrine
*Formerly IBM Research GmbH*
Rüschlikon, Switzerland
*Lumiphase AG*
Zürich, Switzerland
ORCID 0000-0002-3528-4758

Bert Jan Offrein
*IBM Research GmbH*
Rüschlikon, Switzerland
ORCID: 0000-0001-6082-0068

*Abstract*— A Ferroelectric Analog Non-Volatile Memory based on a $WO_x$ electrode and ferroelectric $HfZrO_4$ layer is fabricated at a low thermal budget (~375˚C), enabling BEOL processes and CMOS integration. The devices show suitable properties for integration in crossbar arrays and neural network inference: analog potentiation/depression with constant field or constant pulse width schemes, cycle to cycle and device to device variation <10%, ON/OFF ratio up to 10 and good linearity. The physical mechanisms behind the resistive switching and conduction mechanisms are discussed.

## I. Introduction

Bio-inspired analog hardware accelerators performing the synaptic function in neuromorphic computing are being developed, based on memristors: trimmable resistances from which the value can vary reversibly upon an external stimulus, in a persistent way. The choice of ferroelectric materials is attractive as the polarization can be changed by the application of an external electric field and is remnant. The growth of epitaxial thin films allowed the demonstration of ferroelectric memristors [1], however, integrating such films on CMOS remains challenging and expensive, requiring flip-chip or wafer bonding techniques. The discovery of ferroelectricity in hafnium oxide enabled the demonstration of ferroelectric memristors in the Front- and Back-End-Of-Line (BEOL) [2], [3] through techniques allowing crystallization in the ferroelectric phase with a low thermal budget [4]. Ferroelectric Tunnel Junctions (FTJs) possess only two terminals and are a desirable approach for crossbar array fabrication. An FTJ consists of a ferroelectric layer separating two different electrodes. Upon polarization reversal, the energy profile and thus the transport probability of a carrier across the junction is modified. Their fabrication is challenging as the ferroelectric layer must be thick enough to stabilize ferroelectricity in multiple configurations, but thin enough to allow electric conduction. In this work, we use an oxide semiconducting electrode, $WO_x$. We demonstrate the fabrication of an FTJ showing analog resistive switching in the BEOL with low thermal budget (~375˚C) and low-cost materials (Hf, Zr, W).

## II. BEOL Compatible Fabrication

An asymmetric Metal (M) / Ferroelectric (FE) / semiconductor (SC) / Metal (M) (Fig. 1) layer stack was deposited by Atomic Layer Deposition at 300-350˚C. The semiconducting layer is $WO_{3-x}$, a metal oxide with *n*-type semi-conducting properties due to the presence of oxygen vacancies. The ferroelectric layer is $HfZrO_4$ (HZO), it is capped by metallic TiN to favor the crystallization of HZO in the ferroelectric phase [5] by a ms-flash lamp annealing (ms-FLA) technique: the sample is heated to a moderate temperature of 375˚C, then a 20 ms long flash of 70 $J/cm^2$ in energy is applied to the surface. The $I_d$-$V_g$ characteristics of P- and N-MOS (130 nm) test transistors were not affected by such treatment. W is sputtered on-top and capacitors are then defined by reactive-ion etching of the top electrode (W/TiN). Here, we demonstrate for the first time the fabrication of HZO, ferroelectric, analog non-volatile memories at a thermal budget of ~375°C. X-Ray Diffraction (Fig. 1) confirm the crystallization in the orthorhombic (o-) or tetragonal (t-) phase of HZO, the absence of monoclinic phase, the polycrystalline nature of the films. Junctions do not require wake-up and Dynamic Hysteresis Measurement (DHM) on a 120 µm diameter junction shows that no breakdown is observed after $10^{10}$ switching cycles with triangular pulses of +/- 2V.

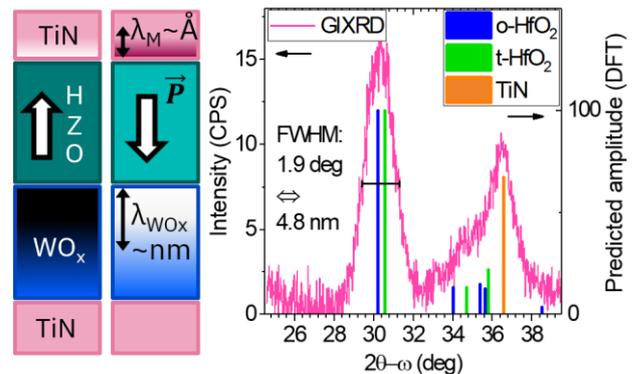

Fig. 1. Left: scheme of the asymmetric screening of HZO ferroelectric polarization by TiN (metal) and $WO_x$ (n-type semiconductor). Right: GIXRD scan showing HZO crystallization in a non-monoclinic phase.

### III. A Ferroelectric Memristor

In this section the resistive memory characteristics of the junction are described. First, a pulse of amplitude $V_{write}$ and duration $t_{width}$ is applied across the junction to align the ferroelectric domains with the applied field. The polarization screening in the M and SC layers occurs over a distance inversely proportional to the carrier density in these two layers, and hence, the energy profile as well as the conductance of the junction is modified. Subsequently, an I(V) sweep in the range +/- 300 mV measures the resistance.

#### A. DC characterization

The device characteristics are highly nonlinear with voltage. The ON/OFF, defined as the ratio of the currents measured in the Low Resistive State, LRS (after applying -1.6 V) and the High Resistive State, HRS (after applying +2.4 V) is >10 for $V_{read}$=100 mV. The LRS, HRS and intermediate states can be reversibly reached and in a remnant way. In the LRS, the polarization points towards the SC (in accumulation mode) and a large coercive field (1.6 MV.cm$^{-1}$) is necessary to switch the ferroelectric domains. In the HRS, the ferroelectric field-effect depletes the SC layer. The poor screening in FE/SC junctions, when the SC is depleted, is usually responsible for the destabilization of the polarization in FTJs. In this work, the metal oxide electrode plays a role in the stabilization of the HRS and allows a large memory window of 1.4 V, as discussed in §IV.A.

#### B. Pulse characterization

Weight update potentiation (depression) is then demonstrated by sending sequences of negative (positive) pulses of 50 µs and increasing amplitude (Fig. 2), showing good repeatability from cycle to cycle.

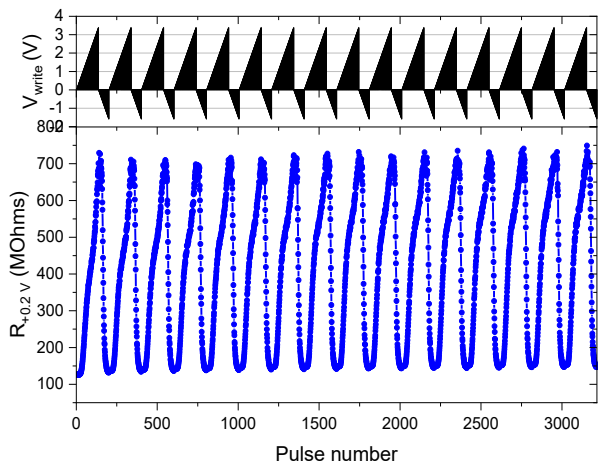

Fig. 2. Read resistance $R^{+0.2V}$ after potentiation/depression pulses of constant duration (50 us) and increasing amplitude. n-SC is grounded.

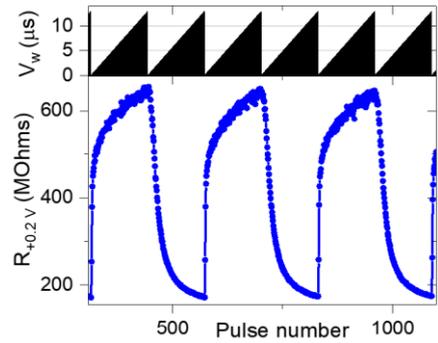

Fig. 3. Read resistance $R^{+0.2V}$ after potentiation/depression pulses of constant field $V_{write}$=3.2 V (-1.4V) and increasing duration $V_w$.

The memristors also demonstrate constant field weight update, by increasing the pulse duration (Fig. 3). Contrary to previous work [2] the ON/OFF is not drastically reduced compared to the increasing amplitude scheme. In Fig. 4 the normalized conductance is fitted by a $\sigma_0(1-e^{-Count/A})$ function were A is the fit parameter [6]. Interestingly linearity is opposite for both schemes (sharp potentiation at constant $V_{ampl}$ vs sharp depression at constant $t_{width}$) showing that symmetry can be tuned using a hybrid scheme by increasing both pulse width and amplitude.

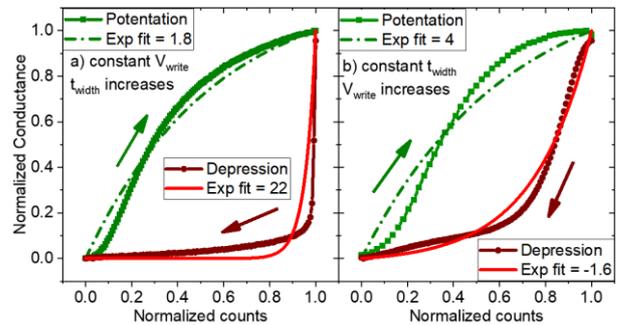

Fig. 4. Normalized conductance vs counts (pulse number). Exp. fit parameters is measured as in ref. [6]. a) for constant field, b) for constant pulse duration.

#### C. Potential integration in a cross-bar array for inference

On top of the low-thermal budget fabrication, the devices show a retention of >10 days and are stable against heating >45°C (Fig. 5). They show a small device to device variation (σ=0.1 in the HRS) and scalability: current density (J) characteristics overlap for capacitors of various sizes. Thanks to the high resistance, the energy of the pulse during writing is < 1 pJ. The non-linearity is high (I(V)/I(V/2)>40 for V>0.5 V) which allows built-in self-selection (limited sneak paths in absence of selectors) in a writing scheme were $V_{write}/2$ is applied to unselected rows [7]. The integration of the junctions in crossbars is limited by their high resistance: extrapolation to sub-micrometric devices leads to currents <pA. This is circumvented by reducing the HZO thickness: in a related work a $10^4$ higher conductance in the HRS is obtained with 4.5 nm thick HZO based junctions.

### IV. Discussion

In order to provide guidelines for the optimization of the ferroelectric memristors for crossbar array fabrication, the physical mechanisms controlling the memory (in particular the stability of the HRS), the conductance and the non-linearity (conduction mechanisms) are explored.

## A. Role of the metal oxide electrode

The stability of the polarization in the HRS and the dependence of the resistive switching on the pulse duration are discussed in terms of oxygen exchange between the ferroelectric and the metal oxide layer. Changing the metal oxide thickness while keeping HZO thickness constant has minor effect on the HRS and LRS, showing that the resistance is dominated by the HZO layer and indicating that the resistive switching is mainly due to ferroelectric switching. However, the dependence of the resistive switching on the pulse duration (Fig. 3) indicates that other than purely ferroelectric effects play a role in the switching mechanism. We anticipate a resistance change through field driven migration of oxygen from the SC to the FE layer, allowing oxygen vacancies in the SC layer able to screen the polarization. This is supported by the observation of a strongly reduced ON/OFF when using $TiO_2$ as SC (allowing little oxygen exchange).

## B. Conduction mechanisms

Devices are tested under temperature cycling (Fig. 5). The ON/OFF is not affected by the temperature, but the resistance decreases upon heating. Such dependence and simulations discard direct tunneling as being the dominant mechanism in these junctions. Log(J/V) is constant at small fields (Ohmic regime, 0~100 mV) and linear with $V^{0.5}$ at higher fields (Poole-Frenkel regime, 200~300 mV). Physical parameters are extracted from the variation of the parameters of the linear regressions with the temperature. In the Poole-Frenkel regime and in the Ohmic regime a small energy barrier of 0.1~0.2 eV is measured, pointing toward conduction via oxygen vacancies in the HZO layer. The oxygen content in HZO is controlled by the growth and annealing conditions and is an additional knob (with the thickness) to tune the conductance of the junctions.

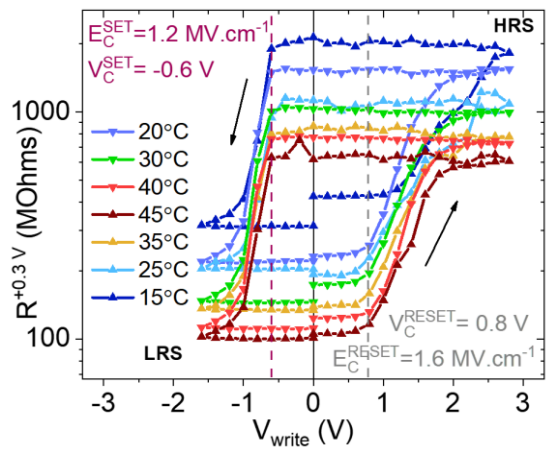

Fig. 5. Read resistance $R^{+0.3V}$ after DC writing $V_{write}$. Very stable HRS ($V_C^{SET}$ = -0.6V) despite screening with depleted SC. R($V_{write}$) loops at increasing and decreasing temperatures. n-SC layer is grounded.

## V. CONCLUSION

The ferroelectric analog non-volatile memory technology presented in this works shows characteristics comparable to other technologies (Table 1). The low-thermal budget process, makes it a promising candidate for the fabrication of crossbar arrays for deep-neural networks accelerators. The devices show analog potentiation/depression with constant field or constant pulse width schemes, which makes them especially interesting for inference where such schemes are easily implemented. This first generation of BEOL, ferroelectric 2-terminals memristors shows non-linearity of (1.9/-4), ON/OFF ratio up to 10 and cycle to cycle and device to device variation <10%. The future optimization of the devices (with thinner HZO layer) is led by the understanding of the memory and conduction mechanisms, with focus on oxygen exchange between the electrode and the ferroelectric layer as well as conduction mediated by oxygen vacancies though the dielectric.

TABLE I. BENCHMARK

| Device type | [8] MO-ECRAM | TaOx/HfOx [9] | PCMO [10] | AlOx/HfO2 [11] | c-FSJ [12] | This work |
|---|---|---|---|---|---|---|
| Non-linearity | <1 | 0.04/-0.63 | 3.7/-6.8 | 1.94/-0.61 | 4.2/-4.2 | 1.9/-4.3 |
| RON | 67KΩ | 100KΩ | 23MΩ | 16.9KΩ | 100KΩ | 100MΩ |
| ON/OFF | 20 | 10 | 6.84 | 4.43 | 21 | 7 |
| Depression | 4V/ 10ns | 1.6V/ 50ns | 2V/ 1ms | 0.9V/ 100µs | 2V/ 80ns | 2.4V/ 50 us |
| Potentiation | -4V/ 10ns | -1.6V/ 50ns | -2V/ 1ms | -1V/ 100µs | -2V/ 80ns | -1.6V/ 50us |
| Cycle-to-cycle var. | <10% | 3.70% | <1% | 5% | <0.5% | 10% |
| Area (µm$^2$) | 40 | 8663.1 | 6292.3 | 21,846 | 184,420 | 14,400 |

ACKNOWLEDGMENT

This work is funded by H2020 "FREEMIND" (840903) and H2020 "BeFerroSynaptic" (871737).

REFERENCES

[1] Garcia, V.; Bibes, M. Ferroelectric Tunnel Junctions for Information Storage and Processing. Nat. Commun. 2014, 5 (1), 4289.

[2] Halter, M.; Bégon-Lours, L.; Bragaglia, V.; Sousa, M.; Offrein, B. J.; Abel, S.; Luisier, M.; Fompeyrine, J. Back-End, CMOS-Compatible Ferroelectric Field-Effect Transistor for Synaptic Weights. ACS Appl. Mater. Interfaces 2020, 12 (15), 17725–17732.

[3] Ni, K.; Smith, et al. S. SoC Logic Compatible Multi-Bit FeMFET Weight Cell for Neuromorphic Applications. 2018 IEEE International Electron Devices Meeting (IEDM); p 13.2.1-13.2.4.

[4] O'Connor, É.; Halter, M.; Eltes, F.; Sousa, M.; Kellock, A.; Abel, S.; Fompeyrine, J. Stabilization of Ferroelectric HfxZr1−xO2 Films Using a Millisecond Flash Lamp Annealing Technique. APL Mater.

[5] Böscke, T. S.; Müller, J.; Bräuhaus, D.; Schröder, U.; Böttger, U. Ferroelectricity in Hafnium Oxide Thin Films. Appl. Phys. Lett. 2011,

[6] Chen, P.; Peng, X.; Yu, S. NeuroSim: A Circuit-Level Macro Model for Benchmarking Neuro-Inspired Architectures in Online Learning. IEEE Trans. CAD. Integr. Circuits Syst. 2018, 37 (12), 3067–3080.

[7] Kim, K.-H.; Gaba, S.; Wheeler, D.; Cruz-Albrecht, J. M.; Hussain, T.; Srinivasa, N.; Lu, W. A Functional Hybrid Memristor Crossbar-Array/CMOS System for Data Storage and Neuromorphic Applications. Nano Lett. 2012, 12 (1), 389–395.

[8] Kim, S. et al. J. Metal-Oxide Based, CMOS-Compatible ECRAM for Deep Learning Accelerator. In 2019 IEEE International Electron Devices Meeting (IEDM); 2019; p 35.7.1-35.7.4.

[9] Huang, X.; Wu, H.; Sekar, D. C.; Nguyen, S. N.; Wang, K.; Qian, H. Optimization of TiN/TaOx/HfO2/TiN RRAM Arrays for Improved Switching and Data Retention. In 2015 IEEE International Memory Workshop (IMW); IEEE: Monterey, CA, USA, 2015; pp 1–4.

[10] S. Park et al. H. Hwang. Neuromorphic Speech Systems Using Advanced ReRAM-Based Synapse. In 2013 IEEE International Electron Devices Meeting; 2013; p 25.6.1-25.6.4.

[11] J. Woo et al. Improved Synaptic Behavior Under Identical Pulses Using AlOx/HfO2 Bilayer RRAM Array for Neuromorphic Systems. IEEE Electron Device Lett. 2016, 37 (8), 994–997.

[12] Si, M. et al. A Novel Scalable Energy-Efficient Synaptic Device: Crossbar Ferroelectric Semiconductor Junction. In 2019 IEEE International Electron Devices Meeting (IEDM); IEEE: San Francisco, CA, USA, 2019; p 6.6.1-6.6.4